\documentclass[a4paper]{spie}  
\usepackage[]{graphicx}
\usepackage{amssymb}
\usepackage{mathrsfs}
\usepackage{gensymb}
\usepackage{color}
\usepackage{url}

\title{The GCT camera for the Cherenkov Telescope Array} 

\author{A.M. Brown\supit{a}, A. Abchiche\supit{b}, D. Allan\supit{a}, J.-P. Amans\supit{c}, T.P. Armstrong\supit{a}, A. Balzer\supit{d}, D.Berge\supit{d}, C. Boisson\supit{c}, J.-J. Bousquet\supit{c}, M. Bryan\supit{d}, G. Buchholtz\supit{b}, P.M. Chadwick\supit{a}, H. Costantini\supit{e}, G.~Cotter\supit{f}, M.K. Daniel\supit{g}, A. De Franco\supit{f}, F. De Frondat\supit{c}, J.-L. Dournaux\supit{c}, D. Dumas\supit{c}, G.~Fasola\supit{c}, S. Funk\supit{h}, J. Gironnet\supit{b},\supit{c}, J.A. Graham\supit{a}, T. Greenshaw\supit{g}, O. Hervet\supit{c}, N. Hidaka\supit{i}, J.A.~Hinton\supit{j}, J.-M. Huet\supit{c}, I. Jegouzo\supit{c}, T. Jogler\supit{h}, M. Kraus\supit{h}, J.S. Lapington\supit{k}, P.~Laporte\supit{c}, J.~Lefaucheur\supit{c}, S. Markoff\supit{d}, T. Melse\supit{c}, L. Mohrmann\supit{h}, P. Molyneux\supit{k}, S.J. Nolan\supit{a}, A.~Okumura\supit{i}, J.P. Osborne\supit{k}, R.D.~Parsons\supit{j}, S. Rosen\supit{k}, D. Ross\supit{k}, G. Rowell\supit{l}, Y. Sato\supit{i}, F.~Sayede\supit{c}, J.~Schmoll\supit{a}, H. Schoorlemmer\supit{j}, M.~Servillat\supit{c}, H.~Sol\supit{c}, V.~Stamatescu\supit{l}, M.~Stephan\supit{d}, R. Stuik\supit{m}, J. Sykes\supit{k}, H. Tajima\supit{i}, J. Thornhill\supit{k}, L. Tibaldo\supit{j}, C. Trichard\supit{e}, J.~Vink\supit{d}, J.J.~Watson\supit{f}, R.~White\supit{j}, N. Yamane\supit{i}, A. Zech\supit{c}, A. Zink\supit{h} and J. Zorn\supit{j} for the CTA Consortium\supit{n}
\skiplinehalf
\supit{a}Department of Physics and Centre for Advanced Instrumentation, University of Durham, South Road, Durham, DH1 3LE, UK; \\
\supit{b}CNRS PSL Research University, Division technique DT-INSU, 1 Place Aristide Briand, 92190 Meudon, France; \\
\supit{c}Observatoire de Paris, CNRS PSL Research University, LUTH and GEPI, Place J. Janssen, 92195, Meudon cedex, France; \\
\supit{d}GRAPPA, University of Amsterdam, Science Park 904, 1098 XH Amsterdam, The Netherlands; \\
\supit{e}Aix Marseille Universite, CNRS/IN2P3, CPPM UMR 7346, 163 avenue de Luminy, case 902, 13288 Marseille, France; \\
\supit{f}Department of Physics, University of Oxford, Keble Road, Oxford OX1 3RH, UK; \\
\supit{g}University of Liverpool, Oliver Lodge Laboratory, P.O. Box 147, Oxford Street, Liverpool L69 3BX, UK; \\
\supit{h}Erlangen Center for Astroparticle Physics (ECAP), Erwin-Rommel-Str. 1, D 91058 Erlangen, Germany; \\
\supit{i}Institute for Space-Earth Environmental Research, Nagoya University, Furo-cho, Chikusa-ku, Nagoya, Aichi 464-8601, Japan; \\
\supit{j}Max-Planck-Institut fur Kernphysik, P.O. Box 103980, D 69029 Heidelberg, Germany; \\
\supit{k}Department of Physics and Astronomy, University of Leicester, University Road, Leicester, LE1 7RH, UK; \\
\supit{l}School of Chemistry and Physics, University of Adelaide, Adelaide 5005, Australia; \\
\supit{m}Leiden Observatory, Leiden University, Postbus 9513, 2300 RA, Leiden, Netherlands; \\
\supit{n}www.cta-observatory.org
}

\authorinfo{Send correspondence to A.M.B: E-mail: anthony.brown@durham.ac.uk, Telephone: $+44$ 191 33 43678}
 
\begin{document} 
  \maketitle 

\begin{abstract}
The Gamma-ray Cherenkov Telescope (GCT) is proposed for the Small-Sized Telescope component of the Cherenkov Telescope Array (CTA). GCT's dual-mirror Schwarzschild-Couder (SC) optical system allows the use of a compact camera with small form-factor photosensors. The GCT camera is $\sim0.4$ m in diameter and has 2048 pixels; each pixel has a $\sim0.2^{\degree}$ angular size, resulting in a wide field-of-view. The design of the GCT camera is high performance at low cost, with the camera housing 32 front-end electronics modules providing full waveform information for all of the camera's 2048 pixels. The first GCT camera prototype, CHEC-M, was commissioned during 2015, culminating in the first Cherenkov images recorded by a SC telescope and the first light of a CTA prototype. In this contribution we give a detailed description of the GCT camera and present preliminary results from CHEC-M's commissioning.  
\end{abstract}



\section{INTRODUCTION}
\label{sec:intro}  

The Cherenkov Telescope Array (CTA) is the next generation ground-based imaging atmospheric Cherenkov telescope (IACT) array and will operate as an open observatory. Building on the strengths of current IACTs, CTA is designed to achieve an order of magnitude improvement in flux sensitivity, with unprecedented energy and angular resolution\cite{cta}. Importantly, CTA will also increase the energy reach of ground-based $\gamma$-ray astronomy, observing photons in the energy range of ~0.02 TeV to beyond 100 TeV. 

To meet these ambitious goals, CTA will consist of three telescope size classes: Small, Medium and Large Sized Telescopes (SST, MST and LST respectively). With a large mirror surface area ($\sim23$ m), the LST is designed to observe the low Cherenkov photon flux associated with $E_{\gamma}<0.2$ TeV photon-induced extended air showers, thus affording CTA with its low energy threshold. With a diameter of $\sim 12$ m, the MSTs will provide the majority of the improvement in flux sensitivity in the $0.1 \leq E_{\gamma} \leq 10$~TeV energy range. The SSTs will extend the high-energy reach of CTA by observing $\gamma$-rays with energies in excess of 5 TeV. For air showers induced by these extremely energetic $\gamma$-rays, the limiting factor is not the number of Cherenkov photons produced in the extended air shower, but rather the number of showers to observe. As such, the SSTs possess the modest primary mirror diameter of $\sim4$ m, a large field-of-view and will be spread over an area greater than $\sim4$~km$^2$; thus maximising the effective area of CTA above $E_{\gamma}>5$ TeV. To achieve all-sky coverage, CTA will consist of two arrays, one in each hemisphere. The baseline northern array will contain 19 telescopes spread over about 1~km$^2$, while the baseline southern array will contain 99 telescopes spread over an area of several square kilometers. About two thirds of the southern array telescopes are envisaged to be SSTs, while no SSTs are currently forseen for the northern array.   

The Gamma-ray Cherenkov Telescope (GCT) is a dual-mirror prototype proposed for the SST component of CTA. The dual-mirror optical design demagnifies the air shower image, allowing a reduction of the telescope's plate scale, and thus enabling small form-factor photosensor elements to be used in the camera. Additionally, the good off-axis point spread function permitted by the optics of the dual-mirror design allows a large field-of-view camera to be built from these photosensors\cite{cameron}. There are two prototypes for the GCT camera: one based on multi-anode photomultipliers (CHEC-M) and one based on silicon photomultipliers (CHEC-S). The first GCT prototype camera (CHEC-M) is currently undergoing detailed commissioning in the laboratory and has undergone initial on-telescope tests at the Paris Observatory, Meudon, during November 2015, on the prototype telescope structure. The culmination of the on-structure commissioning is the recording of the first Cherenkov images from a Schwarzschild-Couder telescope, and the first Cherenkov images recorded by a CTA prototype. In this contribution we give a detailed description of the GCT camera and present preliminary results from the on-sky CHEC-M tests.

\section{CHEC-M Architecture}

\begin{figure}
   \begin{center}
   \begin{tabular}{c}
   \includegraphics[height=8cm]{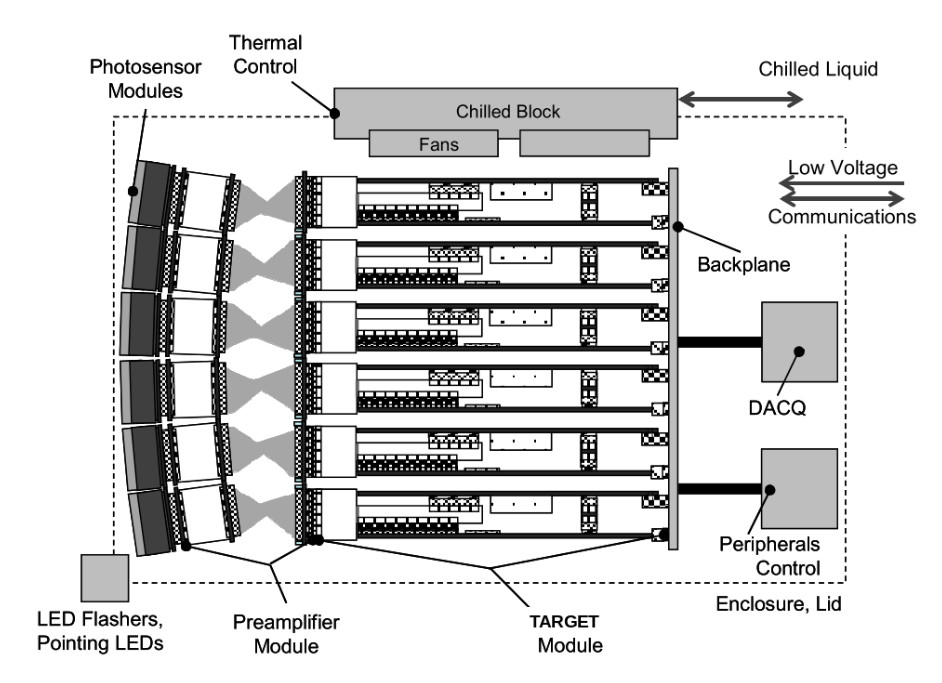}
   \end{tabular}
   \end{center}
   \caption[example] 
   { \label{cad2} Side view showing the internal architecture of CHEC-M. A single photosensor, preamplifier module and TARGET module form a FEE module as discussed in \textsection2.1, while the backplane and 2 DACQ boards form the camera's BEE (see \textsection2.2). \textcopyright The GCT subconsortium.}
   \end{figure}

\begin{figure}
   \begin{center}
   \begin{tabular}{c}
   \includegraphics[height=12cm]{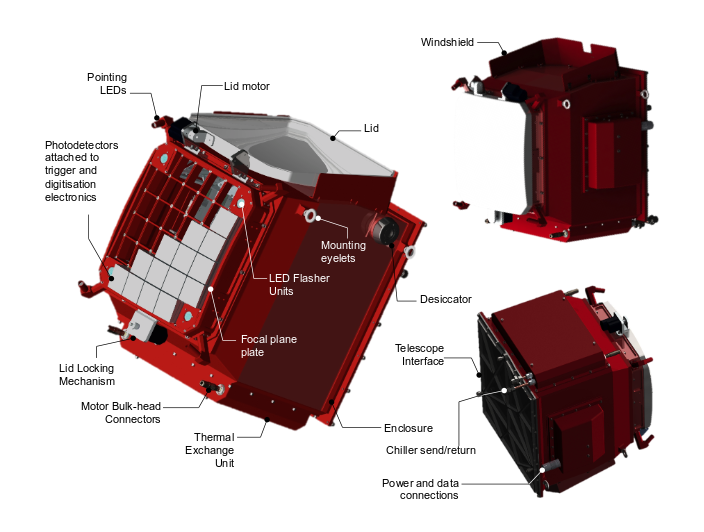}
   \end{tabular}
   \end{center}
   \caption[example] 
   { \label{cad} An overview of the external architecture of CHEC-M. \textcopyright The GCT subconsortium.}
   \end{figure}

An overview of CHEC-M's internal architecture can be seen in Figure \ref{cad2}, with the external architecture shown in Figure \ref{cad}. Taking advantage of the small plate scale associated with the dual-mirror optical system, the total diameter of the GCT camera is 0.4 m in diameter and mass of $\sim45$~kg. The small size and mass allows for easy mounting$/$unmounting of the camera on the GCT telescope structure with no specialist lifting equipment required\cite{gate}. The design philosophy of the GCT camera is high performance at low cost\footnote{Here low cost is on the order of 100 euro per pixel.}, with the camera housing 32 front-end electronics (FEE) modules providing full waveform information for all the camera's 2048 pixels. Described in more detail in \textsection2.1 \& \textsection2.2, each of the 32 FEE modules consists of a photosensor, a preamplifier module and a digitisation module, which are inserted into the camera body through the focal plane plate and pulled into the backplane with a screw on a mechanical standoff at the end of the FEE module. At the corners of the GCT camera focal plane are four LED flasher units. Housing ten UV-peaked LEDs each, these flasher units illuminate the camera focal plane by reflecting off the GCT's secondary mirror, with a range of illumination levels (from 0.1 to 1000 photoelectrons) \cite{me}.

The GCT camera body is a sealed system. As such, to remove the $\sim450$~W of heat generated by the FEE and back-end electronics components of GCT, a chiller system is mounted on the telescope structure, with liquid coolant being circulated via two dry connectors on the camera to an internal chiller block within the camera body. Fans within the camera body circulate air with the aid of a set of mechanical baffles. Since the camera body is designed to be a sealed system, it has undergone environmental testing to confirm the suitability of the design for exposure to the elements. These tests include impact testing, water ingress testing, temperature cycling (from $-25^{\degree}$C to $+40^{\degree}$C), UV exposure and wind tunnel testing. The camera passed all impact, temperature cycling and UV exposure tests. While the camera lid was found to operate in wind speeds up to a maximum of 50~km$/$h, a small amount of water ingress around the lid seal led to a design revision of the lid seal (see \textsection4).    

\subsection{Front-End Electronics}

\begin{figure}
   \begin{center}
   \begin{tabular}{c}
   \includegraphics[height=12cm]{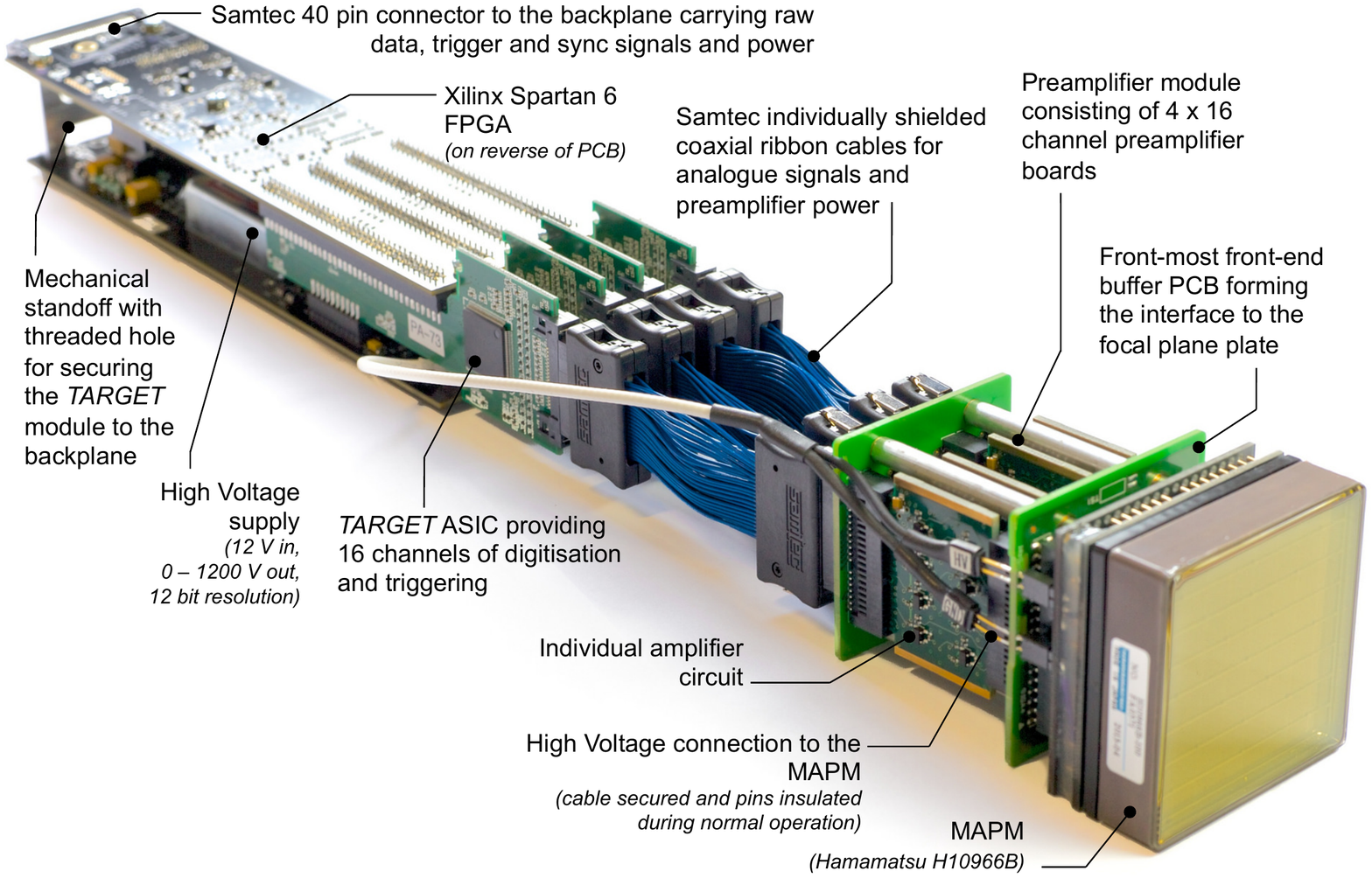}
   \end{tabular}
   \end{center}
   \caption{ \label{fee} Image of one of the FEE modules used in CHEC-M. \textcopyright The GCT subconsortium.}
   \end{figure}

Both of the CHEC-M and CHEC-S prototypes consist of 32 front-end electronics (FEE) modules. An example of a CHEC-M FEE module can be seen in Figure \ref{fee}. There are three primary components of the FEE modules: the photosensor, the preamplifier and the digitisation modules. Starting at the front of a FEE module, the photosensor is connected to a preamplifier module containing 4 PCBs, with each PCB possessing 16 channels. The role of the preamplifier is to shape the analogue signal from each photosensor pixel and stretch it for digitisation. The shaped pulses are then passed from the preamplifier to a digitisation module via shielded coaxial ribbon cables. The flexible ribbon cables allow for the curvature of the focal plane to be removed and also provides the power for the preamplifier module. Possessing four TARGET application specific integrated circuits (ASICs), the digitisation module, referred to as a TARGET module, allows for high temporal resolution digitisation of the signal waveform at a low-cost per pixel\cite{target}. Each TARGET ASIC reads out 16 analogue signals at a rate of 1 GSa/s from the preamplifier in parallel and stores them in seperate analogue buffers. When required, the stored analogue signals are digitised, with 12-bit resolution, for each read out window. The timestep of the read out window is programmable, with 96 ns windows expected for normal camera operation. 

The CHEC-M FEE module uses the TARGET-5 ASIC, which has a dynamic range of $\sim1.2$~V, a read out noise of $\sim0.5$~mV ($0.13$ photoelectrons) and a minimum trigger threshold of $\sim4.5$~mV. At nominal operation voltage the TARGET-5 ASIC saturates at $\sim500$ photoelectrons, though information from waveform fitting extends the TARGET-5's dynamic range beyond this saturation level. 

In addition to signal buffer and digitisation, the TARGET module is also responsible for the first level camera trigger based on the sum of the signal from four neighbouring pixels. This first level trigger is processed on the TARGET module by a Xilinx Spartan-6 FPGA, before being sent to the camera's back-end electronics (BEE). In parallel to the role of signal sampling and digisation, the TARGET module also provides high voltage to the photosensors; for the multi-anode photomultiplier tubes (MaPM) of CHEC-M, this is expected to be in the region of 900-1100 V. At the end of the TARGET module is a mechanical standoff and a 40-pin Samtec connector, through which raw data, power, trigger and synchronisation signals are passed to the BEE.

\subsection{Back-End Electronics}
 
The back-end electronics (BEE) consists of a single backplane PCB and two data acquisition (DACQ) boards. All 32 FEE modules attach directly to the backplane via their mechanical standoffs and 40 pin connector. The backplane houses two FPGAs: one for house-keeping, the other to implement a camera-wide trigger. The house-keeping FPGA controls the supply of electrical power to the TARGET modules and the camera's slow control duties. The trigger FPGA takes all first level triggers from all individual FEE modules to create a camera-wide trigger. Once the trigger FPGA determines a valid camera-wide trigger, all TARGET modules are readout with a unique event number being stamped on the raw data. The backplane then routes all raw data lines to the DACQ boards via User Datagram Protocol (UDP). 

CHEC-M's DACQ boards are custom built White Rabbit\cite{wr} switches connected to the backplane via a Samtec ribbon cable, capable of a transfer rate of $16 \times 1$ Gbps. Communication to the camera, along with all signal and monitoring data from the camera, is via four optical fibres capable of 1~Gbps each, with two fibres being connected to each DACQ board. 

\section{CHEC-M Commissioning}

\begin{figure}
   \begin{center}
   \begin{tabular}{c}
   \includegraphics[height=8cm]{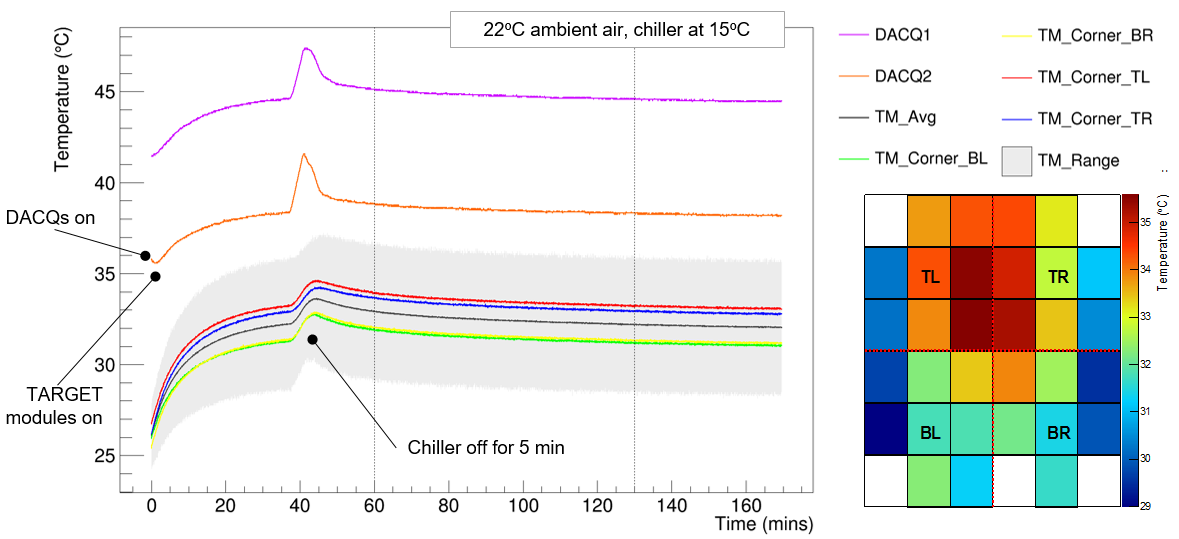}
   \end{tabular}
   \end{center}
   \caption[example] 
   { \label{temp} An example of the interal CHEC-M temperatures during operation over a 170 minute period. \textit{Left}: the temperature of both DACQ boards, and four selected TARGET modules (the positions of which are shown in the right hand insert). \textit{Right}: As view from the focal plane, the temperature for the individual TARGET boards is shown by the colour scale. \textcopyright The GCT subconsortium.}
   \end{figure}

The CHEC-M prototype was assembled in Leicester in mid-2015, and underwent inital commissioning in the lab\cite{andrea}. This initial lab-based commissioning included temperature monitoring \& control, arrival time calibration from pico-second laser flashes, single-photoelectron measurements of every pixel at an operating voltage of 1100V, gain-matching of all FEE modules and the determination of transfer-functions for all cells in the TARGET module's four ASICs. An example of the temperature monitoring tests and the effect of the chiller can be seen in Figure \ref{temp}. Once the lab-based commissioning confirmed the functionality of all FEE and BEE components, and the camera cooling system was found to be sufficient for nominal operation, the CHEC-M prototype was shipped to Paris Observatory, Meudon, for installation on the GCT prototype telescope structure. 

\begin{figure}
   \begin{center}
   \begin{tabular}{c}
   \includegraphics[height=8cm]{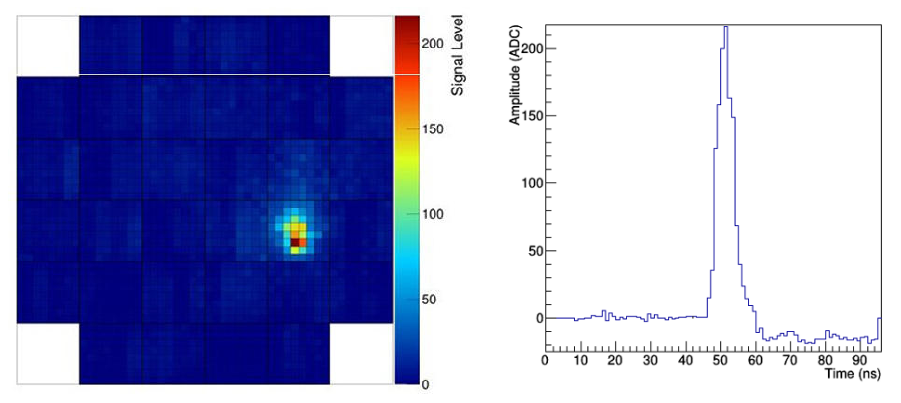}
   \end{tabular}
   \end{center}
   \caption[example] 
   { \label{fl} An example of an air shower event observed during the on-sky tests. \textit{Left}: Uncalibrated camera image of the event, with the colour scale showing the waveform amplitude peak for each pixel in units of ADC. \textit{Right}: 96 ns long waveform of the brightest pixel in the air shower image. \textcopyright The GCT subconsortium.}
   \end{figure}

Integration of the camera with the telescope structure took place during November 2015, with on-sky commissioning occuring during a brief period of clear skies on the 26th of November. While the telescope was pointed away from the centre of Paris, the combination of Parisean lights and rising moon resulted in a large night sky background (NSB) rate $>500$ MHz. This high NSB rate required the MaPMs to be operated at a reduced high voltage of 750~V, which resulted in a higher energy threshold for the observed air showers compared to nominal operation. During the on-sky observations, only one third of the primary mirrors were on the telescope structure, and neither the primary or secondary mirror had undergone mirror alignment. Nonetheless, during the brief window of clear weather, a total of 12 extended air shower events were observed. An example event can be seen in the left plot of Figure \ref{fl}. The colour scale of the event image in Figure \ref{fl} is maximum pixel intensity in units of ADC. It should be noted that due to the low MaPM operating voltage during the on-sky observations, only an estimation of the MaPM gain was possible for the air shower events since the lab-commissioning tests from which gain levels were derived, had been at nominal HV levels. 

The right plot of Figure \ref{fl} is the 96 nanosecond waveform for the brightest pixel in the air shower event. The individual pixel waveform information recorded by the FEE affords CHEC-M a powerful tool with which to analyse events. For example, the waveform shown in Figure \ref{fl} has a temporal profile expected from an extended air shower. Furthermore, the waveform information from all the individual pixels of the event image in Figure \ref{fl} allowed us to confirm the correlated arrival of the Cherenkov photons compared to fluctuations from the bright NSB. As such, these 12 events represent the first Cherenkov images from a Schwarzschild-Couder telescope and the first light of a CTA prototype. 

\section{CHEC-S Architecture}

In parallel to the indepth lab-based tests of CHEC-M that are currently ongoing, the second prototype GCT camera, CHEC-S, is currently being built. The general architecture of both the CHEC-M and CHEC-S prototype is similar, where the fundamental difference between the two prototypes is the composition of the FEE modules. The CHEC-S prototype will use silicon photomultipliers (SiPMs) as the photosensor. Changing from MaPMs to SiPMs will require a revision of the preamplifier PCBs and pulse shaping scheme. Furthermore, building on the experience of CHEC-M, the digitisation board for CHEC-S will be upgraded to 3 board layout, which is envisaged to be populated with the newer TARGET ASICs. This new TARGET board has separate ASICs for triggering and sampling, which have a dynamic range, readout noise and trigger threshold similiar to the TARGET-5 board, whilst be able to perform first-level triggers on the TARGET module. The addition of a third PCB for the newer TARGET module physically separates the triggering, sampling and HV functionality and allows for the physical dimensions of CHEC-S to be compatible with both the GCT and ASTRI\cite{astri} prototype telescopes. The CHEC-S prototype will also see an upgrade to the BEE, with the current two DACQ board configuration with $4 \times 1$ Gbps connection being replaced by a single DACQ board, with a single connection capable of 10 Gbps. 

\begin{figure}
   \begin{center}
   \begin{tabular}{c}
   \includegraphics[height=8cm]{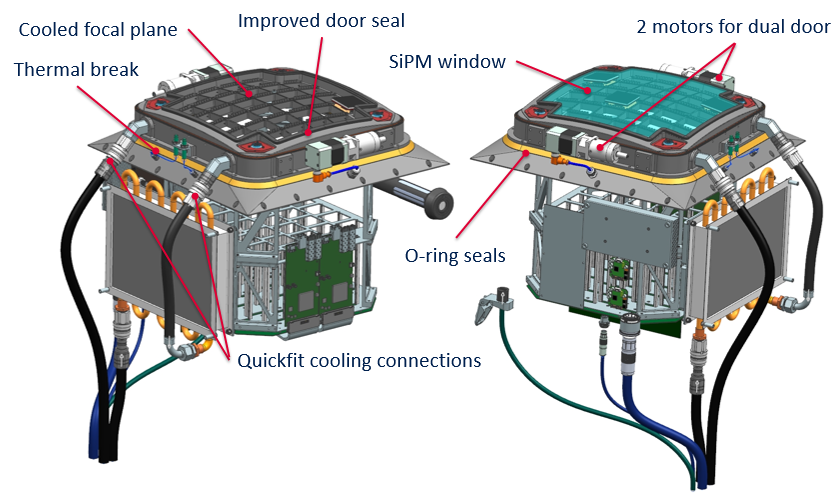}
   \end{tabular}
   \end{center}
   \caption[example] 
   { \label{checsmech} Image showing the key mechanical differences envisaged for the CHEC-S prototype. The instrument enclosure has been omitted to provide visual detail. \textcopyright The GCT subconsortium.}
   \end{figure}

The use of SiPMs also has some implications for the mechanics of the CHEC-S prototype. The gain of SiPMs is temperature dependent which necessitates active temperature cooling of the camera's focal plane plate on which the SiPMs sit. A thermal analysis of CHEC-S's focal plane has shown that cooling of the focal plane plate's edge and ribs is required. As such, CHEC-S is envisaged to have an additional cooling loop from the internal camera cooling block to the focal plane plate (see Figure \ref{checsmech}). A thermal break between the focal plane and the camera body will thermally isolate the focal plane from the rest of the camera.

Building on the lessons learnt from the CHEC-M environmental tests, CHEC-S will also see a redesign of the camera lid. In particular, the single large lid will be replaced by a dual door with a motor controlling each half (see Figure \ref{checsmech}). The dual door design will also have an improved door seal. In addition, between the camera lid and the SiPMs will be a window to further protect the surface of the photosensors. 

\section{Summary}

The GCT camera is a compact, light weight device, with 32 front-end electronics modules providing full waveform information for all of the camera's 2048 pixels. After initial lab-tests confirmed functionality, the CHEC-M prototype was installed on the GCT prototype telescope structure at Paris Observatory, Meudon, for on-sky commissioning. During a brief window of clear weather, the CHEC-M prototype observed 12 extended air shower events. The waveforms of the individual pixels for each of these events confirmed the correlated arrival of the Cherenkov photons when compared to fluctuations from the bright Parisean night sky background, confirming the air shower origin of the observed events. These 12 events represent the first Cherenkov images from a Schwarzschild-Couder telescope and the first light of a CTA prototype. 

After modifications to the telescope structure, and more indepth lab-based commissioning, it is envisaged that a more extended on-sky commissioning campaign will occur at Meudon in late 2016. In parallel to these tests, the CHEC-S prototype is currently being constructed. The architecture of CHEC-S will be similar to CHEC-M, with small changes to allow for the different photosensor technology and the lessons learnt from CHEC-M. Nonetheless, both CHEC-M and CHEC-S will be compatible with both SC SST telescopes currently being prototyped for CTA. 

\acknowledgments We gratefully acknowledge support from the agencies and organizations listed under Funding Agencies at  \url{http://www.cta-observatory.org}.


\end{document}